\documentclass[12pt]{article}
\usepackage{amssymb}
\usepackage{graphicx}
\usepackage{subfigure}
\usepackage[ruled]{algorithm}
\usepackage{algorithmic}

\newtheorem{lemma}{Lemma}
\newtheorem{theorem}{Theorem}
\newtheorem{corollary}{Corollary}

\newcommand{\qed}{\hfill $\Box$}

\def\cH{\mathcal H}

\begin{document}

\title{Optimal Prefix Codes with Fewer Distinct Codeword Lengths are Faster to Construct\thanks{A preliminary version of the paper appeared in the 23rd Annual Symposium on Theoretical Aspects of Computer Science (STACS) 2006 \cite{be2}.}}

\author{Ahmed Belal and Amr Elmasry \thanks{Department of Computer and Systems Engineering, Alexandria University, Egypt}}

\date{}
\maketitle

\begin{abstract}
A new method for constructing minimum-redundancy binary prefix codes is described. Our method does not explicitly build a Huffman tree; instead it uses a property of optimal prefix codes to compute the codeword lengths corresponding to the input weights. 
Let $n$ be the number of weights and $k$ be the number of distinct codeword lengths as produced by the algorithm for the optimum codes. The running time of our algorithm is $O(k \cdot n)$.
Following our previous work in \cite{be}, no algorithm can possibly construct optimal prefix codes in $o(k \cdot n)$ time. 
When the given weights are presorted our algorithm performs $O(9^k \cdot \log^{2k}{n})$ comparisons.
\end{abstract}

\section{Introduction} \label{introduction}

Minimum-redundancy coding plays an important role in data compression applications \cite{zm} as it gives the best possible compression of a finite text when using one static codeword per alphabet symbol. This encoding method is extensively used in various areas of Computer Science like picture compression, data transmission, etc. In accordance, the algorithms used for calculating minimum-redundancy prefix codes that correspond to sets of input symbol weights are of great interest \cite{b,mpl,mk,mt}. 

The minimum-redundancy prefix code problem is to determine, for a given list $W=[w_1,\dots,w_n]$ of $n$ positive symbol weights, a list $L=[l_1,\dots,l_n]$ of $n$ corresponding integer codeword lengths such that $\sum_{i=1}^n 2^{-l_i} \leq 1$,
 and $\sum_{i=1}^n w_i l_i$ is minimized. 
(Throughout the paper, when we state that the Kraft inequality $\sum_{i=1}^n 2^{-l_i} \leq 1$ is satisfied, it is implicitly satisfied with an equality, i.e. $\sum_{i=1}^n 2^{-l_i} = 1$.) Once we have the codeword lengths for a given list of symbol weights, constructing a corresponding prefix code can be easily done in linear time using standard techniques. 

Finding a minimum-redundancy code for $W=[w_1,\dots,w_n]$ is equivalent to finding a binary tree with minimum-weight external path length $\sum_{i=1}^n w(x_i) l(x_i)$ among all binary trees with leaves $x_1,\dots,x_n$, where $w(x_i)=w_i$, and $l(x_i)=l_i$ is the depth of $x_i$ in the corresponding tree. Hence, if we consider a leaf as a weighted node, the minimum-redundancy prefix code problem can be defined as the problem of constructing such an optimal binary tree for a given set of weighted leaves. 

Based on a greedy approach, Huffman's algorithm \cite{h} constructs specific optimal trees, which are referred to as Huffman trees. While every Huffman code is an optimal prefix code, the converse is not true.
The Huffman algorithm starts with a forest $\cH$ of $n$ single-node trees whose values correspond to the given $n$ weights. In the general step, the algorithm selects two roots with the smallest values to become children of a new root, which is added to $\cH$. This new node is assigned a value equals to the sum of the values of its two children. The general step is repeated until there is only one tree in $\cH$, that is called the Huffman tree. The internal nodes of a Huffman tree are thereby assigned values throughout the algorithm; the value of an internal node is the sum of the weights of the leaves of its subtree. With an efficient implementation, this algorithm requires $O(n \log{n})$ time and linear space. The algorithm can be implemented in linear time if the input list was presorted \cite{vl}.

A distribution-sensitive algorithm is an algorithm whose performance relies on how the distribution of the input affects the output. 
For example, a related such algorithm is that of Moffat and Turpin \cite{mt}, where they show how to construct an optimal prefix code on an alphabet of $n$ symbols initially sorted by weight, and including $r$ distinct symbol weights, in $O(r+r\log(n/r))$ time. An output-sensitive algorithm is an algorithm whose performance relies on properties of its output. 
The algorithms proposed in \cite{mpl} are in a sense output sensitive, since their additional space complexities depend on the maximum codeword length $l$ of the output code; the B-LazyHuff algorithm \cite{mpl} runs in $O(n)$ time and requires $O(l)$ extra storage to construct an optimal prefix code on an $n$-symbol alphabet initially sorted by weight.

In this paper we give an output-sensitive recursive algorithm for constructing minimum-redundancy prefix codes;
our algorithm's performance depends on $k$, the number of different codeword lengths 
(that is the number of levels that have leaves in the corresponding optimal binary tree).
As for Huffman's algorithm, we use the RAM model to analyse our algorithm, where we allow comparisons and additions as unit cost operations. We distinguish two cases: the so called \emph{sorted case}, if the sequence of input weights is presorted, and the \emph{unsorted case}, otherwise. For the unsorted case, the running time of our algorithm is $O(k \cdot n)$, which asymptotically surpasses Huffman's bound when $k = o(\log n)$. For the sorted case, our algorithm performs $O(9^k \cdot \log^{2k}{n})$ comparisons, which is sub-linear for sufficiently small $k$. 

Throughout the paper, we interchangeably use the terms leaves and weights. 
Unless otherwise stated, we assume that the input weights are unsorted. Unless explicitly distinguished, a node of a tree can be either a leaf or an internal node. The levels of the tree are numbered bottom up starting from $0$, i.e.~the root has the highest number $l$, its children are at level $l-1$, and the farthest leaves from the root are at level $0$; the length of the codeword assigned to a weight at level $j$ is $l-j$.

The paper is organized as follows. In Section $2$ we recall a property of optimal prefix-code trees on which our construction algorithm relies. In Section $3$ we give the basic algorithm and prove its correctness. We show in Section $4$ how to implement the basic algorithm of Section $3$ to ensure the output-sensitive behavior. An enhancement to the algorithm to achieve the $O(k \cdot n)$ time bound is illustrated in Section $5$. We conclude the paper in Section $6$.

\section{The exclusion property}
  
Consider a binary tree $T$ whose leaves correspond to a list of weights,
such that the value of an internal node of $T$ equals the sum of the weights of the leaves of its subtree.
Assume that the nodes of $T$ are numbered, starting from $1$,  in a non-decreasing order by their values. 
We may assume that the values are distinct, as otherwise we may use a lexicographic order to distinguish between equal values.
The {\it sibling property}, introduced by Gallager \cite{g,v}, states that a tree that corresponds to a prefix code is a Huffman tree if and only if the nodes numbered $2i-1$ and $2i$ are siblings for all $i \geq 1$. 
The sibling property can be alternatively formulated as:

\begin{enumerate}
\item The nodes at each level of $T$ can be numbered in non-decreasing order of their values as $y_1,y_2,y_3\dots$, such that $y_{2i-1}$ and $y_{2i}$ are siblings for all $i \geq 1$.
\item The values of the nodes at a level are not smaller than the values of the nodes at the lower levels.
\end{enumerate}

In \cite{be}, we called this second property the {\it exclusion property}.
In general, building a Huffman tree $T$ that has the sibling property (both the first and second properties) for a list of $n$ weights by evaluating all its internal nodes requires $\Omega(n \log{n})$. This follows from the fact that knowing the values of the internal nodes of $T$ implies knowing the sorted order of the input weights; a problem that requires $\Omega(n \log{n})$ in the comparison-based decision-tree model.
Since it is enough to know which weights will go to which level without necessarily knowing the order of these weights, our main idea is that we do not have to explicitly construct $T$ in order to find the optimal codeword lengths. Instead, we only need to find the values of some of---and not all---the internal nodes while maintaining the exclusion property. 

\section{The basic construction method}

Given a list of weights, we build a corresponding optimal tree level by level in a bottom-up manner. Starting with the lowest level (level $0$), a weight is momentarily assigned to a level as long as its value is less than the sum of two nodes with the currently smallest value(s) at that level; this ensures the exclusion property. Kraft's inequality is enforced by making sure that, at the end of the algorithm, the number of nodes at every level is even, and that the number of nodes at the highest level containing leaves is a power of two. As a result, some weights may be moved upwards from their initially-assigned levels to the higher levels. The details follow.

\subsection{An illustrative example}
To introduce the main ideas, we start with an example. Consider a list with thirty weights: ten weights of value $2$, ten of value $3$, five of value $5$, and five of value $9$. 

\begin{figure}
\centering
	\subfigure[Initial assignment for levels $0$ and $1$]
	{
		\includegraphics[width=0.9\textwidth]{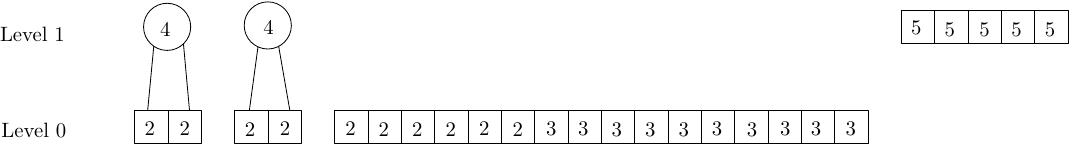}
		\label{fig:e1-1}
	}
	\\
	\vspace{.3in}
	\subfigure[Moving the node with the largest rank at level $1$ to level $2$]
	{
		\includegraphics[width=0.9\textwidth]{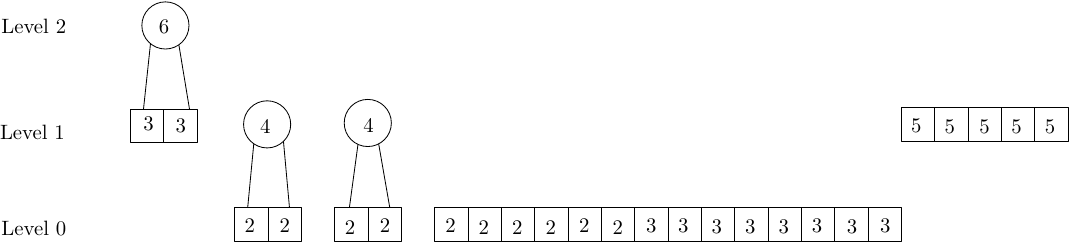}
		\label{fig:e1-2}
	}
	\\
	\vspace{.3in}
	\subfigure[Initial assignment for level $2$]
	{
		\includegraphics[width=0.9\textwidth]{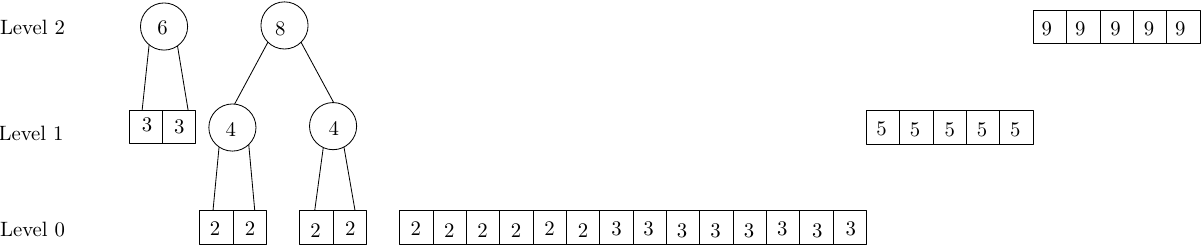}
		\label{fig:e1-3}
	}
	\\
	\vspace{.3in}
	\subfigure[The final weight assignments]
	{
		\includegraphics[width=0.9\textwidth]{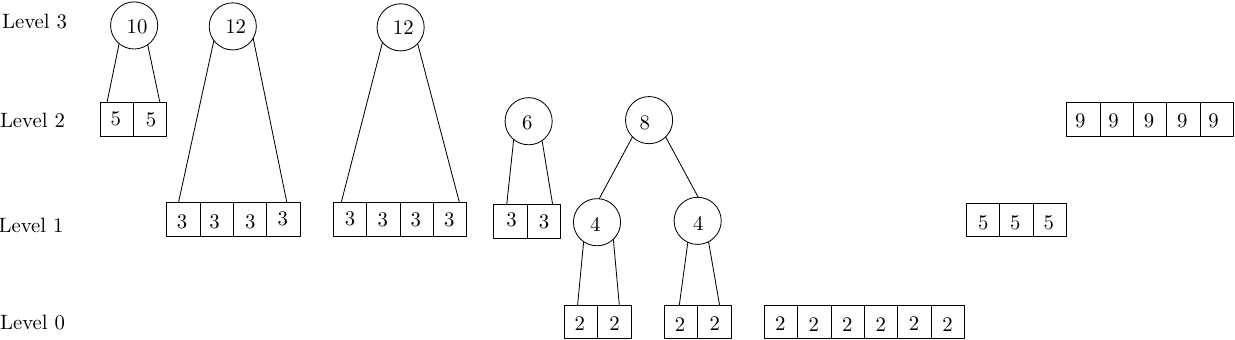}
		\label{fig:e1-4}
	}
\caption{An example for the basic construction method. Squares represent leaves and circles represent internal nodes. The values are written inside the nodes.}
\label{fig:example}
\end{figure}

To construct the optimal codes, we start by finding the smallest two weights in the list; these have the values $2, 2$. We now identify all the weights in the list with value less than $4$, the sum of these two smallest weights. There are twenty such weights: ten weights of value $2$ and ten of value $3$. All these weights are momentarily assigned to the bottom level, that is level $0$ with respect to our level numbering. The number of nodes at level $0$ is now even; so, we go to the next upper level (level $1$). 
We identify the smallest two nodes at level $1$, amongst the two smallest internal nodes resulting from combining nodes already at level $0$ (these have values $4, 4$) and the two smallest weights among those remaining in the list (these have values $5, 5$). It follows that the smallest two nodes at level $1$ will be the two internal nodes $4, 4$ whose sum is $8$. 
All the remaining weights with value less than $8$ are to be momentarily assigned to level $1$. Accordingly, level $1$ now contains an odd number of nodes: ten internal nodes and five weights of value $5$. See Figure~\ref{fig:e1-1}.

To make the number of nodes at level $1$ even, we move the node with the largest value at level $1$ to the, still empty, next upper level (level $2$). The node to be moved, in this case, is an internal node with value $6$. Moving an internal node one level up implies moving the weights in its subtree one level up. So, the subtree consisting of the two weights of value $3$ is moved one level up. At the end of this stage, level $0$ contains ten weights of value $2$ and eight weights of value $3$; level $1$ contains two weights of value $3$ and five weights of value $5$. 
See Figure~\ref{fig:e1-2}.

The currently smallest two internal nodes at level $2$ have values $6, 8$ and the smallest weight in the list has value $9$. This means that all the five remaining weights in the list will be momentarily assigned to level $2$. Now, level $2$ contains eight internal nodes and five weights, for a total of thirteen nodes. See Figure~\ref{fig:e1-3}.

Since we are done with all the weights, we only need to enforce the condition that the number of nodes at level $3$ is a power of two.
All we need to do is to move the three nodes with the largest value(s), from level $2$, one level up. The largest three nodes at level $2$ are the three internal nodes of values $12, 12$ and $10$. So, we move eight weights of value $3$ and two weights of value $5$ one level up. As a result, the number of nodes at level $3$ will be $8$; that is a power of two. The root will then be at level $6$. 
The final distribution of weights will be: ten weights of value $2$ at level $0$; ten weights of value $3$ and three weights of value $5$ at level $1$; and the remaining weights, two of value $5$ and five of value $9$, at level $2$. The corresponding codeword lengths are $6$, $5$ and $4$ respectively. See Figure~\ref{fig:e1-4}.

Note that we have not included all the internal nodes within Figure \ref{fig:example}. We have consciously only drawn the internal nodes that are required to be evaluated by our algorithm; this will be elaborated throughout the next subsections.

\subsection{The algorithm}
The idea of the algorithm should be clear. We construct an optimal tree by maintaining the exclusion property for all the levels. Once the weights are placed in such a way that the exclusion property is satisfied, the property will as well be satisfied among the internal nodes. Adjusting the number of nodes at each level to satisfy the Kraft inequality will not affect the exclusion property, since we are always moving the largest nodes from one level to the next higher level. A formal description follows.
(Note that the main ideas of our basic algorithm described in this subsection are pretty similar to those of the Lazy-Traversal algorithm described in \cite{mpl}.)

\begin{enumerate}

\item Consider a list of input symbol weights $W$ (not necessarily sorted). Two weights with the smallest value(s) are found, removed from $W$, and placed at the lowest level $0$; their sum $S$ is computed. The list $W$ is scanned and all weights less than $S$ are removed from $W$ and placed at level $0$. 
Set $\eta \leftarrow 0$.

\item Repeat the following steps until $W$ is empty: 
\begin{enumerate}
\item If the number of nodes at level $\eta$ is odd, move the subtree rooted at a node with the largest value from level $\eta$ to level $\eta+1$.
\item Determine the new weights that will go to level $\eta+1$ as follows. Find two internal nodes with the smallest value(s) at level $\eta+1$, and the smallest two weights among those remaining in $W$. Find the two smallest values amongst these four, and let their sum be $S$. Scan $W$ for all weights less than $S$, and assign them to level $\eta+1$. 
\item $\eta \longleftarrow \eta + 1$.
\end{enumerate}

\item Set $\hat{\eta} \leftarrow \eta$, i.e. $\hat{\eta}$ is the highest level that is currently assigned weights. Let $m$ be the current number of nodes at level $\hat{\eta}$. If $m$ is not a power of $2$, move $2^{\lceil \lg{m} \rceil} -m$ subtrees rooted at the nodes with the largest value(s) from level $\hat{\eta}$ to level $\hat{\eta}+1$.

\end{enumerate}

\subsection{Proof of correctness}
\label{ex}
To guarantee its optimality, we need to show that both the Kraft inequality and the exclusion property hold for the constructed tree.

\paragraph{\bf Maintaining the Kraft inequality.}
First, we show by induction that the number of nodes at every level, other than the root, is even. 
Assume that this is true up to level $\eta -1$. Since any subtree of our tree is a full binary tree (every node has zero or two children), the number of nodes per level within any subtree is even except for its root. At step $2(a)$ of the algorithm, if the number of nodes at level $\eta$ is odd, we move a subtree one level up. We are thus moving even numbers of nodes between levels among the lower $\eta-1$ levels. Hence, the number of nodes per level remains even among those levels. On the other hand, the number of nodes at level $\eta$ either decreases by $1$ (if the promoted subtree root has no children) or increases by $1$ (if it has two children). Either way, the number of nodes at level $\eta$ becomes even. 

Second, we show that the number of nodes at the last level that is assigned weights is a power of two.
At step $3$ of the algorithm, if $m$ is a power of two, no subtrees are moved up and the Kraft inequality holds. Otherwise, we move $2^{\lceil \lg{m} \rceil} -m$ nodes from level $\hat{\eta}$ to level $\hat{\eta}+1$, leaving $2m - 2^{\lceil \lg{m} \rceil}$ nodes at level $\hat{\eta}$ other than the children of the roots that have just been moved to level $\hat{\eta}+1$. Now, the number of nodes at level $\hat{\eta}+1$ is $m - 2^{\lceil \lg{m} \rceil - 1}$ internal nodes resulting from combining pairs of nodes from level $\hat{\eta}$, plus the $2^{\lceil \lg{m} \rceil} -m$ nodes that we have just moved. This sums up to $2^{\lceil \lg{m} \rceil - 1}$ nodes; that is a power of two.

\paragraph{\bf Maintaining the exclusion property.}
We prove by induction that the---even stronger---sibling property holds, as long as we are evaluating the prescribed internal nodes following Huffman's rules. 
Obviously, the property is not broken when assigning the weights at level 0.
Assume that the sibling property holds up to level $\eta-1$. 
Throughout the algorithm, we maintain the exclusion property by making sure that the sum of two nodes with the smallest value(s) at a level is larger than the values of the nodes at this level. 
When we move a subtree from level $\eta-1$ one level up, its root has the largest value at its level. 
The validity of the sibling property at the lowest $\eta-1$ levels implies that the children of a node with the largest value at a level have the largest value(s) among the nodes at their level. Hence, all the nodes moved up must have had the largest value(s) among the nodes of their level. After being moved one level up, the values of these nodes are thus larger than those of the nodes at the lower levels, and the exclusion property is still maintained.

\subsection{Discussion}

We are still far from being done yet. Though we have introduced the main idea behind the algorithm, some crucial details are still missing.
Namely, we did not show how to evaluate the essential internal nodes. 
Should we have to evaluate all the internal nodes, breaking the $n \log n$ bound would have been impossible \cite{be}.
Fortunately, we only need to evaluate few internal nodes per level. More precisely, except for the last level that is assigned weights, 
we may need to evaluate three internal nodes per level: two with the smallest value(s) and one with the largest value. 
The challenge is how to do that efficiently.
Another pitfall of our basic method, the way we have just described, is that we are to evaluate internal nodes for every level of the tree up to the last level that is assigned weights. The work would thus be proportional to the difference between the length of the maximum and the minimum codeword lengths. We still need to do better, and the way out is to skip doing work at the levels that will not be assigned any weights. Again, the challenge is how to do that efficiently. Subsequently, we explain how to resolve these issues in the next section.

\section{The detailed construction method}
\label{detailed}

Up to this point, we have not shown how to evaluate the internal nodes needed by our basic algorithm, and how to search within the list $W$ to decide which weights are to be assigned to which levels. The main intuition behind the novelty of our approach is that it does not require evaluating all the internal nodes of the tree corresponding to the prefix code, and would thus surpass the $n \log{n}$ bound for several cases. In this section, we show how to implement the basic algorithm in an output-sensitive behavior, filling in the missing details.

\subsection{An illustrative example}
The main idea is clarified through an example with $3n/2 + 2$ weights, where $n$ is a power of two. Assume that the resulting optimal tree will turn out to have $k=3$: $n$ leaves at level $0$, $n/2$ at level $1$, and two at level $\lg{n}$. Note that the $3n/2$ leaves at levels $0$ and $1$ combine to produce two internal nodes at level $\lg{n}$.
It is straightforward to come up with a set of weights that fulfill this outcome. However, to illustrate how the algorithm works for any such set of weights, it is better to handle the situation without explicitly deciding the weights.

For such case, we show how to apply our algorithm so that the optimal codeword lengths are produced in linear time, even if the weights were not presorted. Determining the weights to be assigned to level $0$ can be easily done by finding the smallest two weights and scanning through the list of weights. To determine the weights to be assigned to level $1$, we need to find the values of the smallest two internal nodes at level $1$; these are respectively the sum of the smallest two pairs of weights.
After this point, let us assume that the algorithm uses an oracle that recommends checking level $\lg n$ next.

\begin{figure}
\centering
\includegraphics[width=0.9\textwidth]{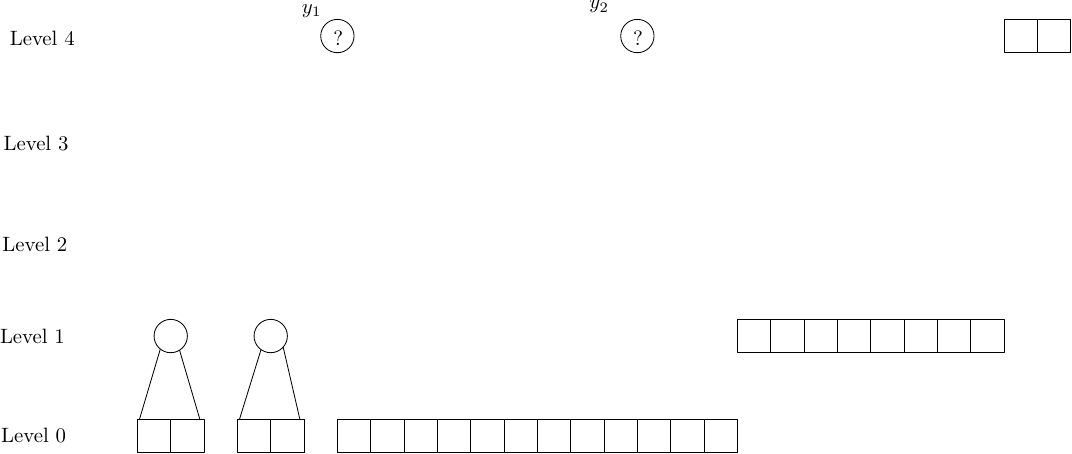}
\caption{An optimal code for $3n/2$ weights, where $n=16$ and $k=3$.}
\label{fig:e2}
\end{figure}

A more involved task is to evaluate the two internal nodes $y_1$ and $y_2$ at level $\lg{n}$, which amounts to identifying the smallest as well as the largest $n/2$ nodes amongst the $n$ nodes at level $1$. The crucial advantage is that we do not need to sort the values of the nodes at level $1$. In addition, we do not need to explicitly evaluate all the $n/2$ internal nodes at level $1$ resulting from the pairwise combinations of the $n$ weights at level $0$. What we really need is the sum of the smaller half as well as the sum of the larger half among the nodes at level $1$. See Figure \ref{fig:e2}. We show next that evaluating $y_1$ and $y_2$ can be done in linear time by a simple pruning procedure. 
 
The nodes at level $1$ consist of two sets; one set has $n/2$ leaves whose weights are known and thus their median $M$ can be found in linear time \cite{cl}, and another set containing $n/2$ internal nodes whose values are not known, but whose median $M'$ can still be computed in linear time by simply finding the two middle weights of the $n$ leaves at level $0$ and adding them. Assuming without loss of generality that $M>M'$, then the larger half of the $n/2$ weights at level $1$ contribute to $y_2$, and the smaller half of the $n$ weights at level $0$ contribute to $y_1$. The above step of finding new medians for the leftovers of the two sets is repeated recursively on a problem half the size. This results in a procedure with a running time that satisfies the recurrence $T(n) = T(n/2) + O(n)$, whose solution results in $T(n)=O(n)$.

If the list of weights was presorted, no comparisons are required to find $M$ or $M'$; we only need to compare them. The total number of comparisons needed satisfies the recurrence $C_s(n) = C_s(n/2) + O(1)$, and hence $C_s(n)=O(\log{n})$.

\subsection{Overview}

Following the basic construction method, the optimal tree will be built bottom up. However, this will not be done level by level; we shall only work in the vicinity of the levels that will end up having leaves. Once we finish assigning weights to a level, we should be able to jump to the next higher level to consider (Section \ref{next-level}). As stated earlier, throughout the algorithm we have to efficiently evaluate some internal nodes. In general, for a specified level and a fixed $t$, we need to be able to evaluate the node with the $t$-th smallest or largest value among the internal nodes, or even among all the nodes, at that level (Section \ref{t-th}). To be able to do that efficiently, we shall start by illustrating a method to evaluate one specific internal node that serves as the median of the nodes on the specified level. More precisely, if one counts the number of weights contributing to each node at the specified level, and supposedly sorts the list of nodes on that level by value, the sought node will be the node that splits this sorted list in two sublists the number of weights contributing to each is as close as possible to the other. We shall show how such a splitting procedure is accomplished in a recursive manner (Section \ref{splitting}).
Once we are able to evaluate such a node, the other tasks---like finding the next level to be assigned weights and finding the $t$-th smallest or largest nodes---can be done by repeatedly calling the splitting procedure. The details are to come up next.

\subsection{The algorithm}
Let $\eta_1 = 0 < \eta_2 < \dots \eta_j$ be the levels that have already been assigned weights after some iterations of our algorithm (other levels only have internal nodes). Let $n_i$ be the number of leaves so far assigned to level $\eta_i$, and $N_{j}=\sum_{i=1}^{j} n_i$. 

At the current iteration, we are looking forward to compute $\eta_{j+1}$, the level that will be next assigned weights by our algorithm. We use the fact that the weights that have already been assigned up to level $\eta_j$ are the only weights that may contribute to the values of the internal nodes below and up to level $\eta_{j+1}$.

Consider the internal node $\chi'_j$ at level $\eta_j$, where the sum of the number of leaves in the subtrees of level-$\eta_j$ internal nodes whose values are smaller than that of $\chi'_j$ is at most but closest to $N_{j-1}/2$. We call $\chi'_j$ the \emph{splitting node of the internal nodes} at level $\eta_j$. In other words, if we define the \emph{multiplicity} of a node to be the number of leaves in its subtree, then $\chi'_j$ is the weighted-by-multiplicity median within the sorted-by-value sequence of the internal nodes at level $\eta_j$. 

Analogously, consider the node $\chi_j$ (not necessarily an internal node) at level $\eta_{j}$, where the sum of the number of leaves in the subtrees of level-$\eta_j$ internal nodes whose values are smaller than that of $\chi_j$ plus the number of level-$\eta_{j}$ leaves whose values are smaller than that of $\chi_j$ is at most but closest to $N_{j}/2$. We call $\chi_j$ the \emph{splitting node of all nodes} at level $\eta_{j}$. Informally, $\chi'_j$ splits the weights below level $\eta_j$ in two groups having almost equal counts, and $\chi_j$ splits the weights below and up to level $\eta_j$ in two groups having almost equal counts.

We extend the notion of splitting nodes to subsets of nodes. 
Let $A$ be a list of weights constituting the leaves of a subset of the internal nodes at level $\eta_j$ having consecutive ranks when sorted by value. 
The splitting node $\chi'_j(A)$ is defined as the weighted-by-multiplicity median within the sorted-by-value sequence of those internal nodes. 
Let $B$ be a subset of leaves at level $\eta_j$ having consecutive ranks when sorted by value. 
The splitting node $\chi_j(A,B)$ is defined as the weighted-by-multiplicity median within the sorted-by-value sequence of $B$ in addition to the internal nodes at level $\eta_j$ whose subtrees have the set of leaves $A$. 

\begin{figure}
\centering
\includegraphics[width=0.9\textwidth]{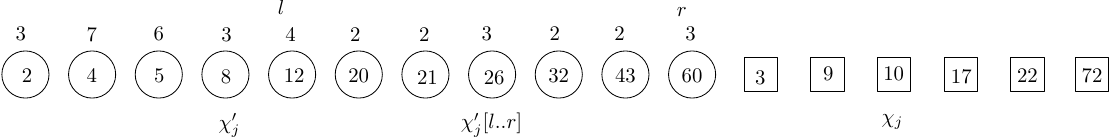}
\caption{Nodes $\chi'_j$, $\chi'_j[l..r]$ and $\chi_j$ at level $\eta_j$.
Numbers inside nodes represent their values, and those outside internal nodes represent their multiplicity. For illustration purposes, the internal nodes are drawn sorted by value and so are the leaves.}
\label{fig:e3}
\end{figure} 

Figure \ref{fig:e3} illustrates those definitions: The internal node with value $8$ is the splitting node $\chi'_j$ of the internal nodes; the sum of the multiplicity of the internal nodes with values smaller than $8$ is $16$, and the sum of the multiplicity of those with values larger than $8$ is $18$.
The leaf with value $10$ is the splitting node $\chi_j$ of all nodes; the sum of the multiplicity of the nodes with values smaller than $10$ is $21$ ($19$ contribute to internal nodes plus $2$ leaves), and the sum of the multiplicity of those with values larger than $10$ is $21$ ($18$ contribute to internal nodes plus $3$ leaves).

\subsubsection{Finding the splitting node}
\label{splitting}
To find the splitting node $\chi_j$ of all nodes at level $\eta_j$, we repeatedly identify splitting nodes of subsets of the internal nodes. The main idea is to apply a pruning procedure in a manner similar to binary search.
Starting with all the leaves and internal nodes at level $\eta_j$, we repeatedly perform the following steps: compare the median of the leaves and the splitting node of the internal nodes, discard sublists of nodes from further consideration, and compute the median of the remaining leaves or the splitting node of the remaining internal nodes. The details follow. 

The method {\it cut}~$(j,\mathcal{L})$  divides the list of weights $\mathcal{L}$ assigned to the $j$ lower levels into two lists, the list of weights $L$ that is assigned to level $\eta_j$ and the list $L'$ of the other weights.
We use the method {\it find-median}($L$) to identify the leaf $M$ with the median weight among the list $L$ of the $n_j$ weights assigned to level $\eta_j$, and partition $L$ into three sublists $\{M\}$, $L_1$ and $L_2$ around $M$, where $L_1$ is the list with the smaller values. We recursively find the splitting node $\chi'_j$ of the internal nodes at level $\eta_{j}$ using the method $find\mbox{-}splitting\mbox{-}internal(j, L')$ given the list $L'$ of the $N_{j-1}$ weights at the levels below $\eta_j$, and partition $L'$ into three sublists $L'_{(\chi'_j)}$, $L'_1$ and $L'_2$ around $\chi'_j$, where $L'_1$ is the list of weights contributing to the internal nodes with the smaller values. We use the method {\it add-weights}($L'_{(\chi'_j)}$) to compute the value of $\chi'_j$ by adding the weights $L'_{(\chi'_j)}$ constituting the leaves of the subtree of $\chi'_j$.

\begin{algorithm}
\footnotesize

\caption{{$find\mbox{-}splitting\mbox{-}all(j,\mathcal{L})$}}
\label{find-split}
\begin{algorithmic}[1]

\STATE $(L, L') \leftarrow$ {\it cut}~$(j,\mathcal{L})$
\STATE $(M, L_1, L_2) \leftarrow find\mbox{-}median(L)$
\IF{$(j=1)$}
	\STATE return ~$|L_1| + 1, \{M\}, L_1, L_2$
\ENDIF
\STATE $(p,L'_{(\chi'_j)}, L'_1, L'_2) \leftarrow find\mbox{-}splitting\mbox{-}internal(j, L')$
\STATE $O_1 \leftarrow O_2 \leftarrow nil$, ~$pos \leftarrow 1$, ~$s_1 \leftarrow \lfloor N_j/2 \rfloor$, ~$s_2 \leftarrow N_j - s_1 - 1$ 
\WHILE{$(|L| \neq 0$ and $|L'| \neq 0)$}
	\IF{($M > add\mbox{-}weights(L'_{(\chi'_j)})$)}
		\IF{$(|L_1| + |L'| - |L'_2| > s_1)$}
			\STATE $O_2 \leftarrow catenate(\{M\}, L_2, O_2)$, ~$s_2 \leftarrow s_2 - 1 - |L_2|$, ~$L \leftarrow L_1$ 	
			\STATE  $(M, L_1, L_2) \leftarrow find\mbox{-}median(L)$ 
		\ELSE	
			\STATE $O_1 \leftarrow catenate(O_1, L'_1, L'_{(\chi'_j)})$, ~$s_1 \leftarrow s_1 - |L'_1| - |L'_{(\chi'_j)}|$, ~$pos \leftarrow pos + p$, ~$L' \leftarrow L'_2$
			\STATE $(p, L'_{(\chi'_j)}, L'_1, L'_2) \leftarrow find\mbox{-}splitting\mbox{-}internal(j, L')$
		\ENDIF
	\ELSE 
		\IF{$(|L_2| + |L'| - |L'_1| > s_2)$}
			\STATE $O_1 \leftarrow catenate(O_1, L_1, \{M\})$, ~$s_1 \leftarrow s_1 - |L_1| - 1$, ~$pos \leftarrow pos + |L_1| + 1$, $L \leftarrow L_2$	
			\STATE  $(M, L_1, L_2) \leftarrow find\mbox{-}median(L)$	 
		\ELSE
			\STATE $O_2 \leftarrow catenate(L'_{(\chi'_j)}, L'_2, O_2)$, ~$s_2 \leftarrow s_2 - |L'_{(\chi'_j)}| - |L'_2|$, ~$L' \leftarrow L'_1$ 	
			\STATE $(p, L'_{(\chi'_j)}, L'_1, L'_2) \leftarrow find\mbox{-}splitting\mbox{-}internal(j, L')$
		\ENDIF
	\ENDIF
\ENDWHILE		
\IF {$(|L| = 0)$} 
	\WHILE{$(|L'_1| > s_1$ or $|L'_2| > s_2)$}
		\IF{$(|L'_2| > s_2)$}
			\STATE $O_1 \leftarrow catenate(O_1, L'_1, L'_{(\chi'_j)})$, ~$s_1 \leftarrow s_1 - |L'_1| - |L'_{(\chi'_j)}|$, ~$pos \leftarrow pos + p$, ~$L' \leftarrow L'_2$
		\ELSE
			\STATE $O_2 \leftarrow catenate(L'_{(\chi'_j)}, L'_2, O_2)$, ~$s_2 \leftarrow s_2 - |L'_{(\chi'_j)}| - |L'_2|$, ~$L' \leftarrow L'_1$
		\ENDIF
		\STATE $(p, L'_{(\chi'_j)}, L'_1, L'_2) \leftarrow find\mbox{-}splitting\mbox{-}internal(j, L')$
	\ENDWHILE	
	\STATE $O_1 \leftarrow catenate(O_1, L'_1)$, ~$O_2 \leftarrow catenate(L'_2, O_2)$, ~$pos \leftarrow pos + |L'_1|$	
	\STATE return ~$pos, L'_{(\chi'_j)}, O_1, O_2$ 	
\ELSE 	
	\WHILE{$(|L_1| > s_1$ or $|L_2| > s_2)$}
		\IF{$(|L_2| > s_2)$}
			\STATE $O_1 \leftarrow catenate(O_1, L_1, \{M\})$, ~$s_1 \leftarrow s_1 - |L_1| - 1$, ~$pos \leftarrow pos + |L_1| + 1$, $L \leftarrow L_2$
		\ELSE
			\STATE $O_2 \leftarrow catenate(\{M\}, L_2, O_2)$, ~$s_2 \leftarrow s_2 - 1 - |L_2|$, ~$L \leftarrow L_1$ 
		\ENDIF
		\STATE $(M, L_1, L_2) \leftarrow find\mbox{-}median(L)$
	\ENDWHILE	
	\STATE $O_1 \leftarrow catenate(O_1, L_1)$, ~$O_2 \leftarrow catenate(L_2, O_2)$, ~$pos \leftarrow pos + |L_1|$
	\STATE return ~$pos, \{M\}, O_1, O_2$ 	
\ENDIF
\end{algorithmic}
\end{algorithm}

Comparing the values of $M$ and $\chi'_j$, assume without loss of generality that $M> \chi'_j$. 
We can then conclude that either the weights in $L_2$ in addition to $M$ must have larger values than $\chi_j$, or the internal nodes corresponding to the weights in $L'_1$ in addition to $\chi'_j$ must have smaller values than $\chi_j$.
In the former case, we set $L$ to $L_1$, and find the new median $M$ and the new lists $L_1$ and $L_2$. In the latter case, we set $L'$ to $L'_2$, and find the new splitting node $\chi'_j$ and the new lists $L'_1$ and $L'_2$. 
The pruning process continues until only the weights contributing to $\chi_j$ remain. As a byproduct, we compute, $O_1$ and $O_2$, the two lists of weights contributing to the nodes at level $\eta_j$ whose values are smaller and respectively larger than $\chi_j$. We also compute, $pos$, the rank of $\chi_j$ among the nodes at level $\eta_j$ when sorted by value.
In the sequel, We use the method $catenate()$ to concatenate the lists in the same order as they appear in its arguments into one list.  See the pseudo-code of Algorithm \ref{find-split}. 

Next, consider the problem of finding the splitting node $\chi'_j$ of the internal nodes at level $\eta_j$. Observe that $\chi_{j-1}$ is a descendant of $\chi'_j$; so, we start by recursively finding the node $\chi_{j-1}$. Let $\alpha$ be the rank of $\chi_{j-1}$ among the nodes at level $\eta_{j-1}$ when sorted by value; the numbering starts from $1$. Knowing that exactly $\lambda = 2^{\eta_j - \eta_{j-1}}$ nodes from level $\eta_{j-1}$ contribute to every internal node at level $\eta_j$, we conclude that the largest $\beta = (\alpha - 1) - \lambda \cdot \lfloor (\alpha - 1) / \lambda \rfloor$ among the $\alpha-1$ nodes whose ranks are smaller than $\chi_{j-1}$ and the smallest $\lambda - \beta - 1$ nodes among those whose ranks are larger than $\chi_{j-1}$ are the nodes contributing to $\chi'_j$. We proceed by finding such nodes; a procedure that requires recursively finding more splitting nodes at level $\eta_{j-1}$ in a way that will be illustrated in the next subsection.
To summarize, the splitting node $\chi'_j$ of level $\eta_j$ is evaluated as follows. The aforementioned pruning procedure is applied to split the weights already assigned to the lower $j-1$ levels to three groups: those contributing to $\chi_{j-1}$, those contributing to the nodes smaller than $\chi_{j-1}$ at level $\eta_{j-1}$, and those contributing to the nodes larger than $\chi_{j-1}$ at level $\eta_{j-1}$. 
The weights contributing to $\chi'_j$ are: the weights of the first group, the weights among the second group contributing to the largest $\beta$ nodes smaller than $\chi_{j-1}$, and the weights among the third group contributing to the smallest $\lambda-\beta-1$ nodes larger than $\chi_{j-1}$. We also compute, $pos$, the rank of $\chi'_j$ among the internal nodes at level $\eta_j$. See the pseudo-code of Algorithm \ref{find-split-internal}.

\begin{algorithm}
\footnotesize
\caption{$find\mbox{-}splitting\mbox{-}internal(j,\mathcal{L})$}
\label{find-split-internal}
\begin{algorithmic}[1]
\STATE $(\alpha, L_{\chi_{j-1}}, O_1, O_2) \leftarrow find\mbox{-}splitting\mbox{-}all(j-1,\mathcal{L})$
\STATE $L_1 \leftarrow L_2 \leftarrow nil$, ~$\lambda \leftarrow 2^{\eta_j - \eta_{j-1}}$
\STATE $\beta \leftarrow \alpha - 1 - \lambda \cdot \lfloor (\alpha - 1) / \lambda \rfloor$
\IF {$(\beta \neq 0)$} 
	\STATE $(O_1, L_1) \leftarrow find\mbox{-}t\mbox{-}largest(\beta, j-1, O_1)$
\ENDIF
\IF {$(\lambda-\beta-1 \neq 0)$} 
	\STATE $(L_2, O_2) \leftarrow find\mbox{-}t\mbox{-}smallest(\lambda-\beta-1, j-1, O_2)$
\ENDIF
\STATE $pos \leftarrow \lceil \alpha/\lambda \rceil$, ~$L'_{(\chi'_j)} \leftarrow catenate(L_1, L_{\chi_{j-1}}, L_2)$
\STATE return ~$pos, L'_{(\chi'_j)}, O_1, O_2$
\end{algorithmic}
\end{algorithm}

\subsubsection{Finding the $t$-th smallest/largest node}
\label{t-th}
Consider the node $\Im$ that has the $t$-th smallest/largest rank among the nodes at level $\eta_{j}$. We use the method {\it find-t-smallest}()/{\it find-t-largest}() to evaluate $\Im$. 

As for the case of finding the splitting node, we find the leaf with the median weight $M$ among the list of the $n_{j}$ weights already assigned to level $\eta_{j}$, and evaluate the splitting node $\chi'_j$ of the internal nodes at level $\eta_{j}$ (applying Algorithm \ref{find-split-internal} recursively) using the list of the $N_{j-1}$ leaves of the lower levels. Comparing $M$ to $\chi'_j$, we discard either $M$ or $\chi'_j$ plus one of the four sublists---the two sublists of $n_{j}$ leaves and the two sublists of $N_{j-1}$ leaves---as not contributing to $\Im$. Repeating this pruning procedure, we identify the weights that contribute to $\Im$ and hence evaluate $\Im$. Accordingly, we also identify the list of weights that contribute to the nodes at level $\eta_{j}$ whose values are smaller and those whose values are larger than $\Im$.

The main ideas of this procedure are pretty similar to those of finding the splitting node, and hence we omit the details and leave them for the reader.

\subsubsection{Computing $\eta_{j+1}$ (the next level that will be assigned weights)} 
\label{next-level}
We start by finding the minimum weight $w$ among the weights still remaining in $W$, 
by applying the method {\it minimum-weight}($W$) that performs a linear scan if the weights are unsorted.
We use the value of $w$ to search within the nodes at level $\eta_{j}$ in a manner similar to binary search. The main idea is to find the maximum number of nodes with the smallest value(s) at level $\eta_{j}$ such that the sum of their values is less than $w$. 
We find the splitting node $\chi_j$ at level $\eta_{j}$, and evaluate the sum of $\chi_j$ plus the weights contributing to the nodes at level $\eta_{j}$ whose values are less than that of $\chi_j$. Comparing this sum with $w$, we decide with which sublist of the $N_{j}$ leaves to proceed to find its splitting node. At the end of this searching procedure, we would have identified the weights contributing to the $\gamma$ smallest nodes at level $\eta_{j}$ such that the sum of their values is less than $w$ and $\gamma$ is maximum. (Note that $\gamma$ is at least $2$, as there are at least two such weights.) We conclude that the level to be considered next for assigning weights is level $\eta_{j}+\lfloor \lg{\gamma} \rfloor$. 
See the pseudo-code of Algorithm \ref{compute-next-level}.

\begin{algorithm}
\footnotesize
\caption{$compute\mbox{-}next\mbox{-}level(j,\mathcal{L},W)$}
\label{compute-next-level}
\begin{algorithmic}[1]
\STATE $w \leftarrow minimum\mbox{-}weight(W)$
\STATE $L \leftarrow \mathcal{L}$, $sum \leftarrow 0$, $\gamma \leftarrow 0$
\WHILE {$(|L| \neq 0)$} 
	\STATE $(\alpha, L_{\chi_j}, O_1, O_2) \leftarrow find\mbox{-}splitting\mbox{-}all(j,L)$
	\STATE $s \leftarrow add\mbox{-}weights(O_1) + add\mbox{-}weights(L_{\chi_j})$  
	\IF {$(sum + s < w)$}
		\STATE $sum \leftarrow sum + s$, ~$\gamma \leftarrow \gamma + \alpha$, ~$L \leftarrow O_2$
	\ELSE
		\STATE $L \leftarrow O_1$
	\ENDIF
\ENDWHILE	
\STATE $\eta_{j+1} \leftarrow \eta_j + \lfloor \lg{\gamma} \rfloor$	
\STATE return ~$\eta_{j+1}$
\end{algorithmic}
\end{algorithm}

To prove the correctness of this procedure, consider any level $\eta$ where $\eta_{j} < \eta < \eta_{j} + \lfloor \lg{\gamma} \rfloor$. The subtrees of two internal nodes with the smallest value(s) at level $\eta$ have at most $2^{\eta - \eta_{j}+1} \leq 2^{\lfloor \lg{\gamma} \rfloor} \leq \gamma$ nodes at level $\eta_{j}$. 
Hence, the sum of the values of such two nodes is less than $w$. For the exclusion property to hold, no weights are to be assigned to any of these levels. On the contrary, the subtrees of the two internal nodes with the smallest values at level $\eta_{j}+\lfloor \lg{\gamma} \rfloor$ have more than $\gamma$ nodes at level $\eta_{j}$, and hence the sum of their values is at least $w$. For the exclusion property to hold, at least the weight $w$ is to be momentarily assigned to level $\eta_{j}+\lfloor \lg{\gamma} \rfloor$.

\subsubsection{Maintaining the Kraft inequality}
After computing the value of $\eta_{j+1}$, we need to maintain the Kraft inequality. This is accomplished by moving the subtrees of some of the nodes with the largest value(s) from level $\eta_{j}$ one level up. Let $n_j$ be the number of nodes currently at level $\eta_{j}$ as counted by the method {\it count-nodes}($j$), and let $\lambda=2^{\eta_{j+1}-\eta_{j}}$. We shall show that the number of subtrees to be moved up is $\nu = \lambda \cdot  \lceil n_j/\lambda \rceil - n_j$. 
See the pseudo-code of Algorithm \ref{maintain-kraft}.
Note that when $\eta_{j+1}-\eta_{j}=1$ (as in the case of our basic algorithm), then $\nu=1$ if $n_j$ is odd and $\nu=0$ otherwise.

\begin{algorithm}
\footnotesize
\caption{$maintain\mbox{-}Kraft\mbox{-}inequality(j, \mathcal{L}, W)$}
\label{maintain-kraft}
\begin{algorithmic}[1]
\STATE $n_j \leftarrow count\mbox{-}nodes(j)$
\IF{$(|W| \neq 0)$}
	\STATE $\lambda \leftarrow 2^{\eta_{j+1} - \eta_j}$
	\STATE $\nu \leftarrow \lambda \cdot  \lceil n_j/\lambda \rceil - n_j$
\ELSE
	\STATE $\nu \leftarrow 2^{\lceil \lg{n_j} \rceil} - n_j$
\ENDIF
\STATE $(\mathcal{L}, L) \leftarrow find\mbox{-}t\mbox{-}largest(\nu,j,\mathcal{L})$
\FOR {every weight $w \in L$}
	\STATE $level(w) \leftarrow level(w) +1$
\ENDFOR
\end{algorithmic}
\end{algorithm}

To establish the correctness of this procedure, we need to show that both the Kraft inequality and the exclusion property hold. For a realizable construction, the number of nodes at level $\eta_{j}$ has to be even, and if $\eta_{j+1}-\eta_{j} \neq 1$, the number of nodes at level $\eta_{j}+1$ has to divide $\lambda/2$.
If $n_j$ divides $\lambda$, no subtrees are moved to level $\eta_{j}+1$ and the Kraft inequality holds. If $n_j$ does not divide $\lambda$, then $\lambda \cdot \lceil n_j /\lambda \rceil - n_j$ nodes are moved to level $\eta_{j}+1$, leaving $2 n_j - \lambda \cdot \lceil n_j / \lambda \rceil$ nodes at level $\eta_{j}$ other than those of the subtrees that have just been moved one level up. Now, the number of nodes at level $\eta_{j}+1$ is $n_j - \lambda \cdot \lceil n_j / \lambda \rceil /2$ internal nodes with children remaining (not moved up) at level $\eta_{j}$, plus the $\lambda \cdot \lceil n_j / \lambda \rceil - n_j$ nodes that we have just moved. This sums up to $\lambda \cdot \lceil n_j / \lambda \rceil /2$ nodes, which divides $\lambda/2$, and the Kraft inequality holds. The exclusion property holds following the same argument given in Section \ref{ex}. 
Kraft's inequality for the highest level that is assigned leaves, i.e. when $|W| = 0$, is correctly maintained also following the argument given in Section \ref{ex}. 

\subsubsection{Summary of the algorithm} 

\begin{enumerate}
\item The smallest two weights are found, moved from $W$ to the lowest level $\eta_1=0$, and their sum $S$ is computed. The rest of $W$ is searched for weights less than $S$, which are assigned to level $0$ as well. Set $j \leftarrow 1$.

\item Repeat the following steps until $W$ is empty:

\begin{enumerate}
\item Compute $\eta_{j+1}$ (the next level that will be assigned weights). 
\item Maintain the Kraft inequality at level $\eta_{j}$ (by moving the $\nu = \lambda \cdot  \lceil n_j/\lambda \rceil - n_j$ subtrees with the largest values from this level one level up, where $\lambda=2^{\eta_{j+1}-\eta_{j}}$ and $n_j$ is the current number of nodes at level $\eta_j$).
\item Find the values of the smallest two internal nodes at level $\eta_{j+1}$, and the smallest two weights from those remaining in $W$. Find two nodes with the smallest value(s) among these four, and let their sum be $S$.
\item Search the rest of $W$, and assign the weights less than $S$ to level $\eta_{j+1}$.
\item $j \longleftarrow j+1$.
\end{enumerate}
\item If $n_j$ is not a power of $2$, move the $2^{\lceil \lg{n_j} \rceil} - n_j$ subtrees rooted at the nodes with the largest values from level $\eta_j$ to level $\eta_j+1$.
\end{enumerate}

\subsection{Complexity analysis}

Let $T(j,|\mathcal{L}|)$ be an upper bound on the time required by our algorithm to find the splitting node (and also for the $t$-th smallest node) of a set of nodes at level $\eta_j$ which are roots of subtrees having the list of leaves $\mathcal{L}$. It follows that $T(j, N_j)$ bounds the time to find $\chi_j$. 
Let $T'(j,|\mathcal{L}|)$ be an upper bound on the time required to find the splitting node of a set of internal nodes at level $\eta_j$ which are roots of subtrees having the list of leaves $\mathcal{L}$. It follows that $T'(j, N_{j-1})$ bounds the time to find $\chi'_j$.
 
First, consider Algorithm \ref{find-split}. The total amount of work  required to find the medians among the $n_{j}$ weights assigned to level $\eta_{j}$ in all the recursive calls is $O(n_{j})$. During the pruning procedure to locate $\chi_{j}$, the time for the $i$-th recursive call to find a splitting node of the internal nodes at level $\eta_{j}$ is at most $T'(j, \lfloor N_{j-1}/2^{i-1} \rfloor)$. The pruning procedure, therefore, requires at most $\sum_{i=1}^{\hat{i}} T'(j, \lfloor N_{j-1}/2^{i-1} \rfloor) + O(n_{j})$ time, where $\hat{i}=\lfloor \lg{N_{j-1} \rfloor}$. 
Mathematically, $T(j, N_{j}) \leq \sum_{i=1}^{\hat{i}} T'(j, \lfloor N_{j-1}/2^{i-1} \rfloor) + O(n_{j})$.

Second, consider Algorithm \ref{find-split-internal}. To find the splitting node of the internal nodes at level $\eta_j$ we find the splitting node of all the nodes at level $\eta_{j-1}$ for the same list of weights. We also find the $t$-th smallest and largest nodes among each half of this list of weights; the time for each of these two calls is at most $T(j-1, \lfloor |\mathcal{L}|/2 \rfloor)$. 
Mathematically, $T'(j, |\mathcal{L}|) \leq  T(j-1, |\mathcal{L}|) + 2 ~T(j-1, \lfloor |\mathcal{L}|/2 \rfloor) + O(1)$.

Summing up the bounds, the next relations follow:
\begin{eqnarray*}
T(1, N_{1}) & = & O(n_1), \\
T(j, 1) & = & O(1), \\
T(j, N_{j}) & \leq & \sum_{i=1}^{\hat{i}} T(j-1, \lfloor N_{j-1}/2^{i-1} \rfloor) + 2 \sum_{i=1}^{\hat{i}-1} T(j-1, \lfloor N_{j-1}/2^{i} \rfloor) + O(n_{j}). 
\end{eqnarray*}

Substitute with $T(a,b) \leq c \cdot 4^a \cdot b$, for $1 \leq a < j$, $1 \leq b < N_{j}$, and some big constant $c$. 
Then, we induce for $j \geq 2$ that 
\begin{eqnarray*}
T(j, N_{j}) & \leq & c \cdot 4^{j-1} \cdot N_{j-1} \Bigl(\sum_{i=1}^{\infty} 1/2^{i-1} + 2 \sum_{i=1}^{\infty} 1/2^{i}\Bigr) +O(n_{j}) \\
& \leq & c \cdot 4^{j} \cdot N_{j-1} + c \cdot n_{j}.
\end{eqnarray*}

Using the fact that $N_{j} = N_{j-1} + n_{j}$, then

\[ T(j, N_{j}) =  O(4^j \cdot N_{j}). \]

Consider the case when the list of weights $W$ is already sorted. 
Let $C_s(j,N_j)$ be the number of comparisons required 
to find the splitting node at level $\eta_j$. The number of comparisons, in all recursive calls, performed against the medians among the $n_{j}$ weights assigned to level $\eta_{j}$, is at most $3 \lg{(n_{j}+1)}$ (at most $\lg{(n_{j}+1)}$ comparisons to find $\chi_{j}$, another $\lg{(n_{j}+1)}$ to find the $\beta$-th largest node among the nodes smaller than $\chi_{j}$, and a third $\lg{(n_{j}+1)}$ to find the $(\lambda-\beta-1)$-th smallest node among those larger than $\chi_{j}$). 
The next relations follow:
\begin{eqnarray*}
C_s(1, N_{1}) & = & 0, \\
C_s(j, 1) & = & 0, \\
C_s(j, N_{j}) & \leq & \sum_{i=1}^{\hat{i}} C_s(j-1, \lfloor N_{j-1}/2^{i-1} \rfloor) + 2 \sum_{i=1}^{\hat{i}-1} C_s(j-1, \lfloor N_{j-1}/2^{i} \rfloor) + 3 \lg{(n_{j}+1)}. 
\end{eqnarray*}

Since the number of terms forming the summands in the previous relation is at most $3 \hat{i} - 1 < 3 \lg {N_{j-1}}$, it follows that 
\begin{eqnarray*}
C_s(j, N_{j}) & < & 3 \lg{N_{j-1}} \cdot C_s(j-1, N_{j-1}) + 3 \lg{(n_{j}+1)} \\
& < & 3 \lg{N_{j}} \cdot C_s(j-1, N_{j-1}) + 3 \lg{N_{j}}.
\end{eqnarray*}

Substitute with $C_s(a,b) \leq 3^{a-1} \cdot \sum_{i=1}^{a-1} \lg^i{b}$, for $1 \leq a < j$, $1 \leq b < N_{j}$. We thus obtain for $j \geq 2$ that 
\begin{eqnarray*}
C_s(j, N_{j}) & < & 3 \lg{N_{j}} \cdot 3^{j-2} \cdot \sum_{i=1}^{j-2} \lg^{i}{N_{j}} + 3 \lg{N_{j}} \\
& \leq & 3^{j-1} \cdot \sum_{i=1}^{j-1} \lg^{i}{N_{j}} \\
& = & O(3^j \cdot \log^{j-1}{N_{j}}).
\end{eqnarray*}

Third, consider Algorithm \ref{compute-next-level}. The time required by this procedure is dominated by the $O(n)$ time to find the minimum weight $w$ among the weights remaining in $W$ plus the time for the calls to find the splitting nodes. Let $T''(j, N_{j})$ be the time required by this procedure, and let $\hat{i} = \lfloor \lg N_j \rfloor$. Then, 

\[T''(j, N_{j}) \leq \sum_{i=1}^{\hat{i}} T(j, \lfloor N_{j}/2^{i-1} \rfloor) + O(n) = O(4^j \cdot N_{j}+n).\]

Let $C''_s(j, N_{j})$ be the number of comparisons required by Algorithm \ref{compute-next-level} when the list of weights $W$ is presorted. Then, 
\[C''_s(j, N_{j}) \leq \sum_{i=1}^{\hat{i}} C_s(j, \lfloor N_{j}/2^{i-1} \rfloor) +O(1) = O(3^j \cdot \log^{j}{N_{j}}).\]

Finally, consider Algorithm \ref{maintain-kraft}. The required time is dominated by the time to find the weights contributing to the $\nu$ nodes with the largest values at level $\eta_{j}$, which is $O(4^j \cdot N_{j})$. 
If $W$ is presorted, the number of comparisons involved is $O(3^j \cdot \log^{j-1}{N_{j}})$.

Using the bounds deduced for the described steps of the detailed algorithm, we conclude that the time required by the general iteration is $O(4^j \cdot N_{j} + n)$. 

If $W$ is presorted, For achieving the claimed bounds, the only point left to be mentioned is how to find the weights of $W$ smaller than the sum of the values of the smallest two nodes at level $\eta_{j}$. Once this sum is evaluated, we apply an exponential search that is followed by a binary search on the weights of $W$; this requires $O(\log{n_j})$ comparisons.
It follows that the required number of comparisons for the general iteration is still $O(3^j \cdot \log^{j}{N_{j}})$.

To complete the analysis we need to show the effect of maintaining the Kraft inequality on the complexity of the algorithm. Consider the scenario when, as a result of moving subtrees one level up, all the weights at a level move up to the next level that already had other weights. As a result, the number of levels that contain leaves decreases. It is possible that within a single iteration the number of the levels that contain leaves decreases to half its value. If this happens for several iterations, the amount of work done by the algorithm would have been significantly larger compared to $k$, the actual number of the final distinct codeword lengths. Fortunately, this scenario will not happen quite often. In the next lemma, we bound the number of iterations performed by the algorithm by $2k$. We even show---the stronger result---that at any step of the algorithm the number of levels that are assigned weights at this point
is at most twice the number of the final distinct codeword lengths for the weights that have been assigned so far.

\begin{lemma}
\label{l2}
Consider the set of weights that will have the $\tau$-th largest codeword length among the optimal prefix codes at the end of the algorithm. During the course of execution of the algorithm these weights will be assigned to at most two consecutive (among the levels that are assigned weights) levels, with level numbers at most $2\tau-1$ and $2\tau$. Hence, the number of iterations performed by the algorithm is at most $2k$, where $k$ is the number of distinct codeword lengths for the optimal prefix codes. 
\end{lemma}

\paragraph{Proof.}

Consider a set of weights that will turn out to have the same final codeword length. Assume (for the sake of contradiction) that, during the course of execution of the algorithm, these weights are initially assigned to three levels $\eta_j < \eta_{j+1} < \eta_{j+2}$. The only way for such weights to later catch each other at the same level would be as a result of moving subtrees up to maintain the Kraft inequality. 

We show next that for this to happen, just after assigning the weights at level $\eta_j$, there would exist an internal node $y$ at level $\eta_j+1$ whose value is strictly smaller than a weight $w$ among those weights that will be assigned to level $\eta_{j+2}$. From the exclusion property, the value of $y$ is at most $w$; so, we need to show that the value of $y$ is not equal to $w$. Let $s_1$ and $s_2$ be the two smallest nodes at level $\eta_{j+1}$ just after the weights have been assigned to level $\eta_{j+1}$. If the value of $y$ is smaller than that of $s_1$ and $s_2$, it is then smaller than $w$ by the exclusion property.
If the value of $y$ is equal to that of either $s_1$ or $s_2$, then the value of $y$ is not equal to $w$; for otherwise, $w$ would have been assigned first to level $\eta_{j+1}$. We are left with the case where the value of $y$ is larger than those of $s_1$ and $s_2$ (this only happens when $\eta_{j+1} = \eta_j+1$). In accordance, $s_1$ and $s_2$ are both leaves; for otherwise, we could have named one of them $y$ and the claim would have been fulfilled.
If the value of $y$ is equal to $w$, then it is equal to the sum of $s_1$ and $s_2$; for otherwise, $w$ would have been assigned first to level $\eta_{j+1}$.
In such a case, $s_1$ and $s_2$ are equal in value and both are equal to the value of both children of $y$.
But, following the way our algorithm works, this can not happen as the algorithm should have then assigned $s_1$ and $s_2$ first to level $\eta_j$ and not $\eta_{j+1}$. We conclude that the value of $y$ is indeed strictly smaller than $w$.

Suppose next that, at some point during the algorithm, the weights that were assigned first to level $\eta_j$ are moved up to catch the weights at level $\eta_{j+2}$. It follows that $y$ will accordingly move to level $\eta_{j+2}+1$. Since the value of $y$ is smaller than $w$, the exclusion property will not hold; a fact that contradicts the behavior of our algorithm.  It follows that these weights were initially assigned to at most two levels.

We prove the second part of the lemma by induction on $\tau$. The base case follows easily for $\tau=1$. Assume that the argument is true for $\tau-1$. By induction, the levels of the weights that will have the $(\tau-1)$-th largest optimal codeword length will be assigned to the at most $2\tau-3$ and $2\tau-2$ levels. From the exclusion property, it follows that the weights that have the $\tau$-th largest optimal codeword length must be at the next upper levels. Using the first part of the lemma, the number of such levels is at most two. It follows that these weights are assigned to the, at most, $2\tau-1$ and $2\tau$ levels among the levels that are assigned weights.

It follows that the weights with the $\tau$-th largest optimal codeword length will be assigned within $2\tau$ iterations. Since the number of distinct codeword lengths is $k$, the number of iterations performed by the algorithm is at most $2k$.
\qed
\medskip

Using Lemma \ref{l2}, the time required by our algorithm to assign the set of weights whose optimal codeword length is the $j$-th largest, among all distinct lengths, is $O(4^{2j} \cdot n)= O(16^j \cdot n)$. Summing for all such lengths, the total time required by our algorithm is $\sum_{j=1}^{k} O(16^j \cdot n) = O(16^k \cdot n)$. 

Consider the case when the list of weights $W$ is presorted. Using Lemma \ref{l2}, the number of comparisons performed to assign the weights whose codeword length is the $j$-th largest among all distinct lengths is $O(9^j \cdot \log^{2j}{n})$. Summing for all such lengths, the number of comparisons performed by our algorithm is $\sum_{j=1}^{k} O(9^j \cdot \log^{2j}{n}) = O(9^k  \cdot \log^{2k}{n})$. The next theorem follows.

\begin{theorem}
Constructing a minimum-redundancy prefix code for a set of $n$ weights presorted by value can be done using $O(9^k \cdot \log^{2k}{n})$ comparisons, where $k$ is the number of distinct codeword lengths of the output code. 
\end{theorem}

\begin{corollary}
If the list of weights was presorted, for $k < c \cdot \lg{n}/\lg{\lg{n^3}}$ and any constant $c < 1/2$, the above algorithm requires $o(n)$ comparisons. 
\end{corollary}

\section{The improved algorithm}
The drawback of the algorithm we described in the previous section is that it uses many recursive median-finding calls. We perform the following enhancement for the unsorted case. The main idea we use here is to incrementally reorder the already assigned weights throughout the algorithm 
while assigning more weights. This is done by partitioning the assigned weights into unsorted blocks, such that the weights of one block are smaller or equal to the weights of the succeeding block. The time bound required by the recursive calls improves when dealing with shorter blocks.

The invariant we maintain is that during the execution of the general iteration of the algorithm, after assigning weights to $j$ levels, the weights that have already been assigned to a level $\eta_{j'}$, $j' \leq j$, are partitioned into blocks each of size at most $\lceil n_{j'}/8^{j-j'} \rceil$, such that the weights of each block are smaller or equal to the weights of the next block. To accomplish this invariant, once we assign weights to a level, the weights of each block among those already assigned to all the lower levels are partitioned into eight almost equal blocks, by finding the weights at the seven quartiles and partitioning around these weights. Each partitioning process takes linear time \cite{cl}.
Using Lemma \ref{l2}, the number of iterations performed by the algorithm is at most $2k$. The amount of work required for this partitioning is $O(n)$ per iterations, for a total of an extra $O(k \cdot n)$ time for the partitioning procedures. 

For $j-j' \geq \log_8{N_{j'}}$, all the weights assigned to level $\eta_{j'}$ and the lower levels are already sorted as a result of the partitioning procedure. We maintain the invariant that the internal nodes of all these levels are evaluated and their values are explicitly stored, by performing the following incremental evaluation procedure once the above condition is satisfied. 
The internal nodes at level $\eta_{j'-1}$ have been evaluated in a previous iteration, since the above invariant must have been fulfilled earlier for level $\eta_{j'-1}$. What we need to do at this iteration is to merge the sorted sequence of the weights assigned to level $\eta_{j'-1}$ with the sorted sequence of the internal nodes of level $\eta_{j'-1}$, and evaluate the corresponding internal nodes at level $\eta_{j'}$. This extra work can be done in a total of $O(n)$ time per iteration. As a consequence, finding the value of a node---the splitting node or the $t$-th smallest or largest node---within any of these lower levels is done in constant time; see the recursive relations below.

The dominant step for all our procedures is to find the median weight among the weights already assigned to a level $\eta_{j'}$. This step can now be done faster. To find such median weight, we can identify the block that has such median (the middle block) in constant time, then we find the required weight in $O(\lceil n_{j'}/8^{j-j'} \rceil)$, which is the size of the block at this level. To perform any of our procedures---like finding the splitting node at a level---the median finding process can be repeated a number of times equals to the logarithm of the number of blocks; that is $O(\log 8^{j-j'} + 1) = O(j-j'+1)$. The total work done at that level besides the recursive calls will then be $O(\lceil n_{j'} \cdot (j-j'+1) /8^{j-j'} \rceil)$.
Let $G(j', j, N_{j'})$ be the time performed by the improved algorithm at and below level $j'$ while assigning the weights at level $j$, where $j' \leq j$ and $\hat{i} = \lfloor \lg N_{j'-1} \rfloor$.
The next recursive relations follow:\\
If $j-j' > \log_8{n_{j'}}$
\begin{eqnarray*}
G(j', j, N_{j'}) & = & 0,  
\end{eqnarray*}
otherwise
\begin{eqnarray*}
G(j', j, 1) & = & O(1), \\
G(1, j, N_{1}) & = & O(\lceil n_1 \cdot j /8^{j-1} \rceil), \\
G(j', j, N_{j'}) & \leq & \sum_{i=1}^{\hat{i}+1} G(j'-1, j, \lfloor N_{j'-1}/2^{i-1} \rfloor) + 2\sum_{i=1}^{\hat{i}}  G(j'-1, j, \lfloor N_{j'-1}/2^{i} \rfloor) \\
                &      & + ~ O(\lceil n_{j'}\cdot (j-j'+1)/8^{j-j'}.
\end{eqnarray*}
Substitute with $G(a,j, b) \leq c \cdot b\cdot (j-a+1) /8^{j-a}$, for $1 \leq a < j'$, $1 \leq b < N_{j'}$, $1 \leq j-a \leq \log_8{b}$, and some big constant $c$. Then, 
\begin{eqnarray*}
G(j', j, N_{j'}) & < & c \cdot N_{j'-1} \cdot(j-j'+2) / 8^{j-j'+1} \cdot \Bigl(\sum_{i=1}^{\infty} 1/2^{i-1} + 2 \sum_{i=1}^{\infty} 1/2^{i}\Bigr) \\
& & + \lceil c \cdot n_{j'}\cdot (j-j'+1) /8^{j-j'} \rceil\\
& \leq &  c \cdot (N_{j'-1} + n_{j'})\cdot (j-j'+1) / 8^{j-j'} .
\end{eqnarray*}
Since $N_{j'} = N_{j'-1} + n_{j'}$, it follows that 
\[G(j', j, N_{j'}) = O(N_{j'} \cdot (j-j'+1) / 8^{j-j'}).\]
The work done to assign the weights at level $j$ is therefore 

\[G(j, j, N_{j}) = O(N_{j}) = O(n).\]

Since the number of iterations performed by the algorithm is at most $2k$, by Lemma \ref{l2}. Summing up for these iterations, the running time for performing the recursive calls is $O(k \cdot n)$. The next main theorem follows.

\begin{theorem}
Constructing a minimum-redundancy prefix code for a set of $n$ unsorted weights can be done in $O(k \cdot n)$ time, where $k$ is the number of distinct codeword lengths of the output code.
\end{theorem}

\section{Comments}
We gave an output-sensitive algorithm for constructing minimum-redundancy prefix codes, whose running time is $O(k \cdot n)$. For sufficiently small values of $k$, this algorithm asymptotically improves over other known algorithms that require $O(n \log{n})$ time. It is quite interesting to know that the construction of optimal prefix codes can be done in linear time when $k$ turns out to be a constant. For sufficiently small values of $k$, if the sequence of weights was presorted, the number of comparisons performed by our algorithm is asymptotically better than other known algorithms that require $O(n)$ comparisons. It is quite interesting to know that the number of comparisons required to construct an optimal prefix code for a sorted sequence of weights is poly-logarithmic when $k$ turns out to be a constant. 

We have shown in \cite{be} that the verification of a given prefix code for optimality requires $\Omega(n \log{n})$ in the algebraic decision-tree model. That lower bound was illustrated through an example of a prefix code with $k=\Theta(\log{n})$ distinct codeword lengths. Since the construction is harder than the verification, constructing the codes for such example thus requires $\Omega(n \log{n})$ time. This implies that there is no algorithm for constructing optimal prefix codes that runs in $o(k \cdot n)$, for otherwise we could have been able to construct optimal codes for the example in \cite{be} in $o(n \log{n})$.

One remaining question is if it is possible or not to make our algorithm faster in practice by avoiding so many recursive calls to a median-finding procedure.

\end{document}